\documentclass[a4paper]{llncs}
\usepackage{graphicx}
\usepackage{amsmath}
\usepackage{hyperref}
\usepackage{color}
\usepackage{subfigure}

\begin{document}

\title{Frustration and collectivity in spatial networks}

\author{Anna Ma\'nka-Kraso\'n and Krzysztof Ku{\l}akowski}

\institute{Faculty of Physics and Applied Computer Science, AGH University of Science and Technology, al. Mickiewicza 30, PL-30059 Krak\'ow, Poland \\
\email{impresja@gmail.com, kulakowski@novell.ftj.agh.edu.pl}
}

\maketitle
\begin{abstract}
In random networks decorated with Ising spins, an increase of the density of frustrations reduces the transition temperature of the spin-glass ordering. This result is in contradiction to the Bethe theory. Here we investigate if this effect depends on the small-world property of the network. The results on the specific heat and the spin susceptibility indicate that the effect appears also in spatial networks. 
\end{abstract}


\section{Introduction}

A random network is an archetypal example of a complex system \cite{dgm}. If we decorate the network nodes with some additional variables,
the problem can be mapped to several applications. In the simplest case, these variables are two-valued; these can be sex or opinion (yes or no)
in social networks, states ON and OFF in genetic networks, 'sell' and 'buy' in trade networks and so on. Information on stationary states
of one such system can be useful for the whole class of problems in various areas of science. The subject of this text is the antiferromagnetic 
network, where the variables are Ising spins $S_i=\pm 1/2$. As it was discussed in \cite{dgm}, the ground state problem of this network can be 
mapped onto the MAX-CUT problem, which belongs to the class of NP-complete optimization problems. Also, the state of the antiferromagnetic 
network in a weak magnetic field gives an information on the minimal vertex cover of the network, which is another famous NP-complete problem \cite{dgm}.
Further, in the ground state of the antiferromagnetic network all neighboring spins should be antiparallel, i.e. their product should be equal to -1.
This can be seen as an equivalent to the problem of satisfiability of $K$ conditions imposed on $N$ variables, where $N$ is the number of nodes and 
$K$ is the number of links. The problem of satisfiability is known also to be NP-complete \cite{gar}. Here we are particularly interested in an 
ifluence of the network topology on the collective magnetic state of the Ising network. The topology is to be characterized by the clustering coefficient $C$, which is a measure of the density of triads of linked nodes in the network. In antiferromagnetic systems, these triads contribute to 
the ground state energy, because three neighboring spins of a triad cannot be antiparallel simultaneously to each other. This effect is known as 
spin frustration. When the frustration is combined with the topological disorder of a random network, the ground state of the system is expected
to be the spin-glass state, a most mysterious magnetic phase which remains under dispute for more than thirty years \cite{by,fh,ns}. These facts suggest
that a search on random network with Ising spins and antiferromagnetic interaction can be worthwhile.\\

Here we are interested in an influence of the density of frustration on the phase transition from the disordered paramagnetic phase to the ordered
spin-glass phase. In our Ising systems, the interaction is short-ranged and the dynamics is ruled by a simple Monte-Carlo heat-bath algorithm, with one parameter $J/k_BT$, i.e. the ratio of the exchange energy constant $J$ to the thermal energy $k_BT$ \cite{her}. Despite this simplicity, the low-temperature phase is very complex and multi-degenerate even in periodic systems, where the topological disorder is absent \cite{gos}. Current theory
of Ising spin-glasses in random networks ignores the contribution of frustrations, reducing the network to a tree \cite{dgm}. In a 'tree-like 
structure' closed loops as triads are absent. In the case of trees the Bethe theory is known to work well \cite{dgm,bax}. In our considerations, the Bethe formula for the transition temperature $T_{SG}$ from the paramagnetic to the spin glass phase \cite{dgm}

\begin{equation}
\frac{-2J}{T_{SG}}=\ln\frac{\sqrt{B}+1}{\sqrt{B}-1}
\end{equation}
serves as a reference case without frustrations. Here $B=z_2/z_1$ is the ratio of the mean number of second neighbours to the mean number of the first neighbours. Then, the transition temperature $T_{SG}$ depends on the network topology. We note that in our network there is no bond disorder; all interactions are antiferromagnetic \cite{task}. \\

In our former texts, we calculated the transition temperature $T_{SG}$ of the Erd\"os-R\'enyi networks \cite{amk1} and of the regular network \cite{amk2}.
The results indicated that on the contrary to the anticipations of the Bethe theory $T_{SG}$ decreases with the clustering coefficient $C$. However,
in both cases we dealt with the networks endowed with the small-world property. It is not clear what dimension should be assigned to these networks,
but it is well known that the dimensionality and in general the network topology influences the values of temperatures of phase transitions \cite{stan,bax,ho}. On the other hand, many real networks are embedded in the three-dimensional space - these are called spatial networks \cite{hbp}. In particular, magnetic systems belong obviously to this class. Therefore, the aim of this work is to calculate the phase transition temperature $T_{SG}$ again for the spatial networks. As in our previous texts \cite{amk1,amk2} the clustering coefficient $C$ is varied as to investigate the influence of the density of frustrations on $T_{SG}$.\\

\begin{figure}[ht] 
\centering
{\centering \resizebox*{10cm}{7cm}{\rotatebox{-90}{\includegraphics{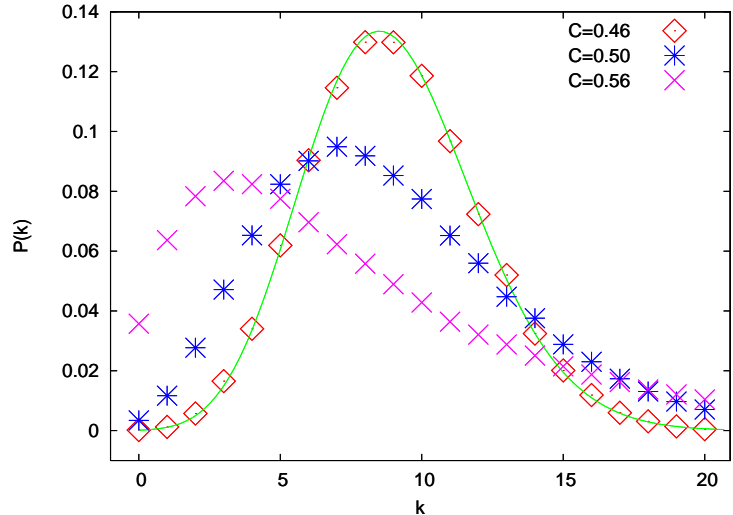}}}} 
\caption{The degree distribution for the mean degree $<k>=9$ and three different values of the clustering coefficient $C$.}
\label{fig-1}
\end{figure}

In the next section we describe the calculation scheme, including the details on the control of the clustering coefficient. Third section is devoted to 
our numerical results. These are the thermal dependences of the magnetic susceptibility $\chi(T)$ and of the spacific heat $C_v(T)$. Final conclusions are given in the last section.

\section{Calculations}

The spatial network is constructed as follows. Three coordinates of the positions of nodes are selected randomly from the homogeneous distribution
between 0 and 1. Two nodes are linked if their mutual distance is not larger than some critical value $a$. In this way $a$ controls the mean number of neighbours, i.e. the mean degree $<k>$ of the network. In networks obtained in this way, the degree distribution $P(k)$ agrees with the Poisson curve.
As a rule, the number of nodes $N=10^5$. Then, to obtain $<k>=4$ and $<k>=9$ we use $a=0.0212$ and $a=0.0278$. The mean degree $<k>$ is found to be proportional to $a^{2.91}$. This departure from the cubic function is due to the open boundary conditions. In two above cases, the values of the clustering coefficient $C$ are respectively 0.42 and 0.47.\\

Now we intend to produce spatial networks with given mean degree $<k>$ and with enhanced clusterization coefficient $C$. This is done in two steps. First
we adjust the radius $a$ to obtain a smaller $<k>$, than desired. Next we apply the procedure proposed by Holme and Kim \cite{hoki}: for each pair 
of neighbours of the same node a link between these neighbours is added with a given probability $p'$. This $p'$ is tuned as to obtain a desired
value of the mean degree $<k>$. Simultaneously, the clustering coefficient $C$ is higher. In this way we obtain $C$ between 0.42 and 0.46 for $<k>=4$,
and between 0.47 and 0.56 for $<k>=9$. The degree distribution $P(k)$ in the network with enhanced $C$ differs from the Poisson distribution, as shown in Fig. 1.\\

The dynamics of the system is ruled by the standard Monte Carlo heat-bath algorithm \cite{her}. We checked that for temperature $T>0.5$, the system
equilibrates after $10^3$ Monte Carlo steps; in one step each spin is checked. Sample runs ensured that after this time, the specific heat $C_v$
calculations from the thermal derivative of energy and from the energy fluctuations give - within the numerical accuracy - the same results.

\begin{figure}[ht] 
\centering
{\centering \resizebox*{10cm}{7cm}{\rotatebox{-90}{\includegraphics{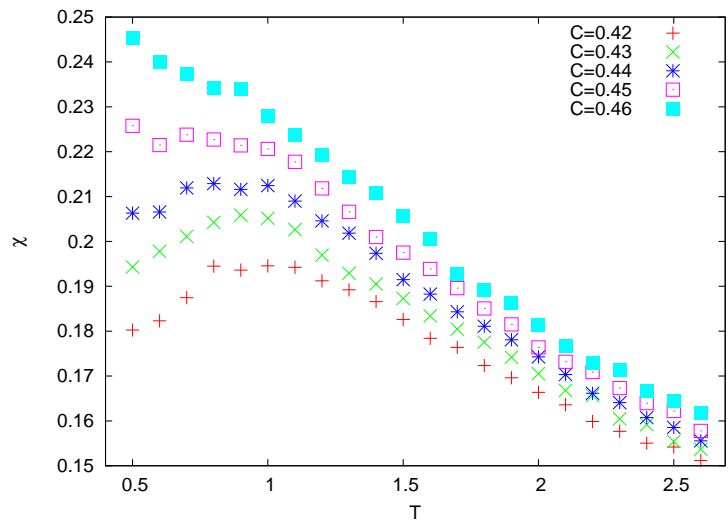}}}} 
\caption{The magnetic susceptibility $\chi(T)$ for $<k>=4$ and different values of the clustering coefficient $C$.}
\label{fig-2}
\end{figure}

\begin{figure}[ht] 
\centering
{\centering \resizebox*{10cm}{7cm}{\rotatebox{-90}{\includegraphics{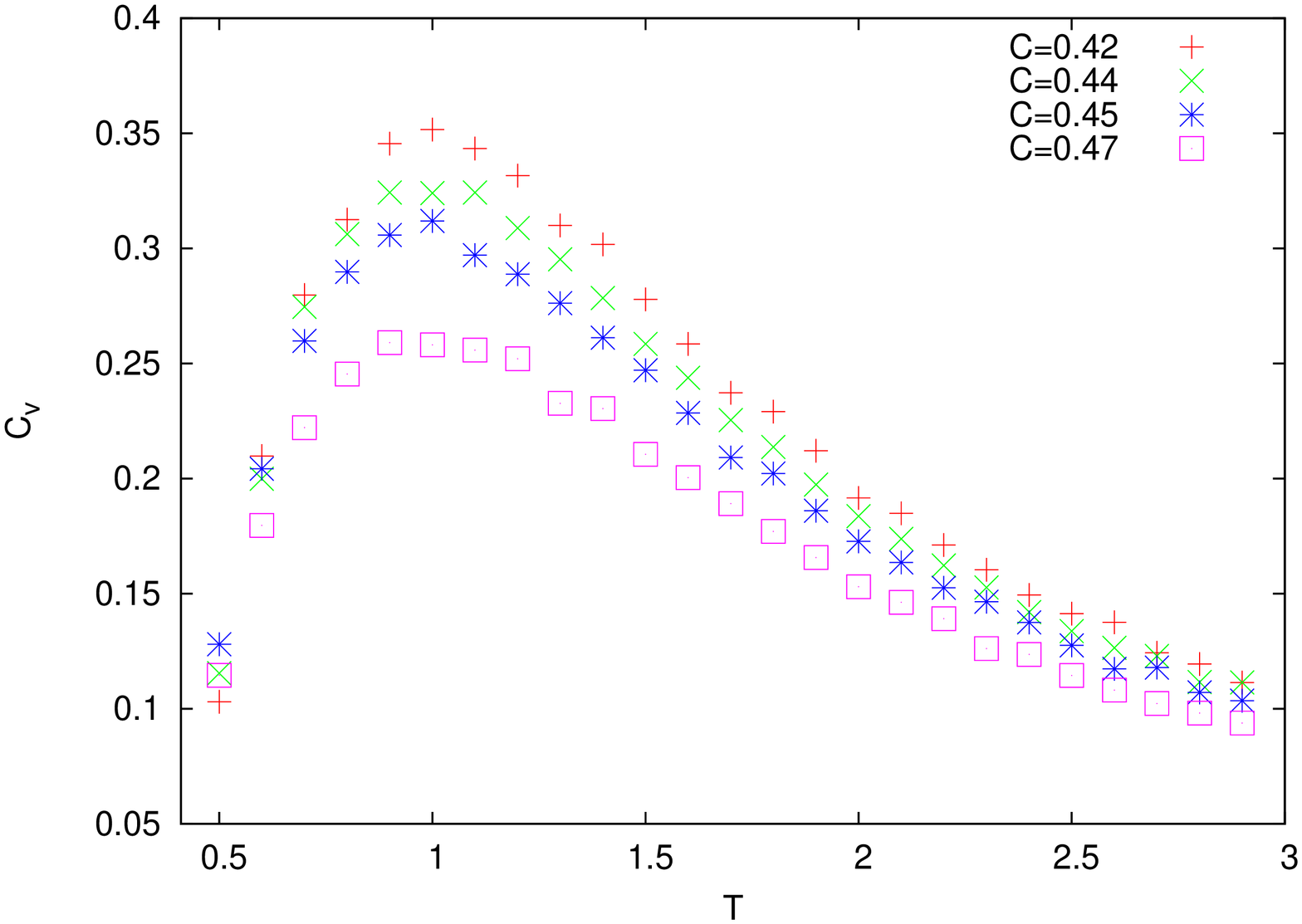}}}} 
\caption{The magnetic specific heat $C_v(T)$ for $<k>=4$ and different values of the clustering coefficient $C$.}
\label{fig-3}
\end{figure}

\begin{figure}[ht] 
\centering
{\centering \resizebox*{10cm}{7cm}{\rotatebox{-90}{\includegraphics{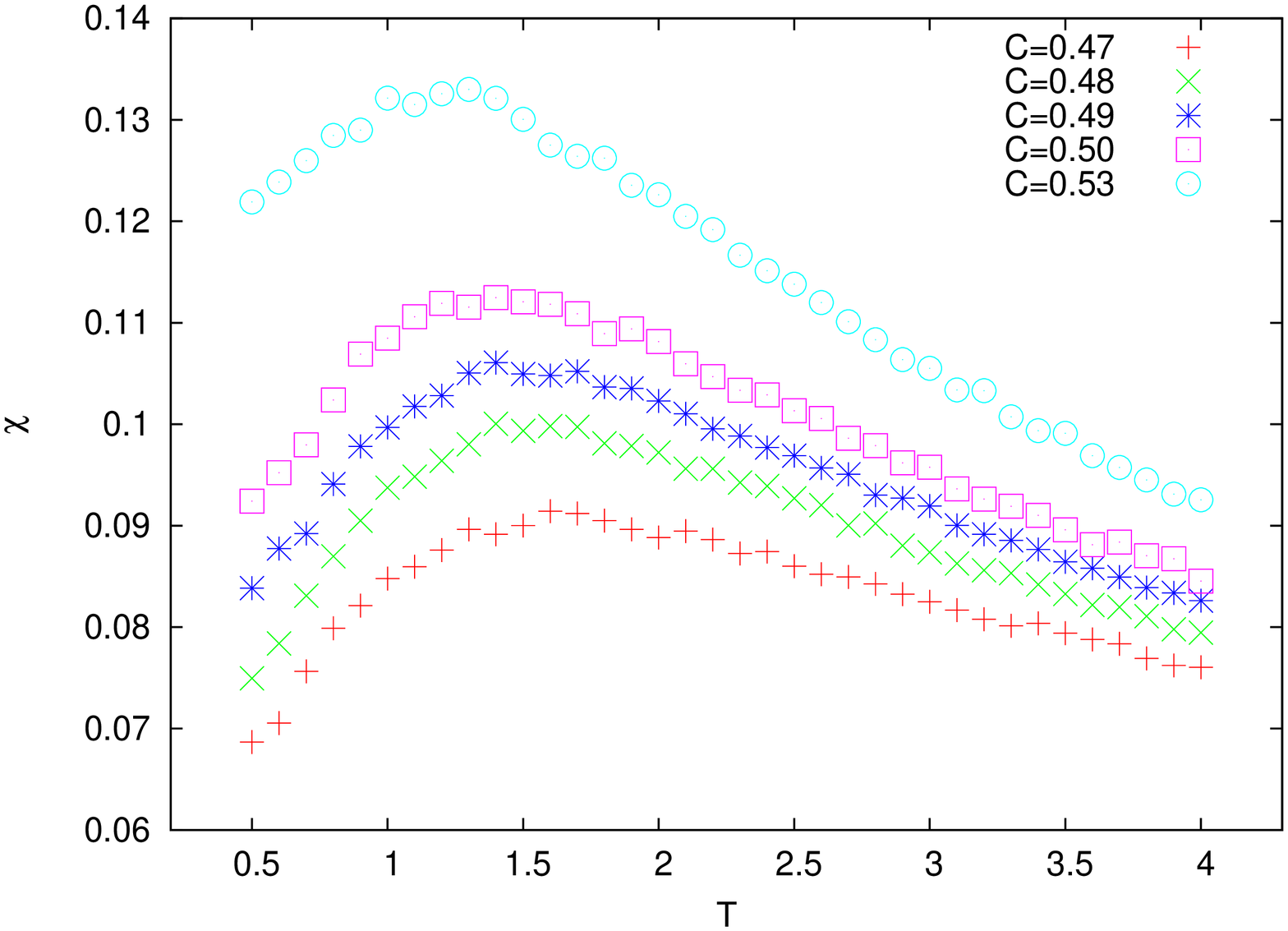}}}} 
\caption{The magnetic susceptibility $\chi(T)$ for $<k>=9$ and different values of the clustering coefficient $C$.}
\label{fig-4}
\end{figure}

\begin{figure}[ht] 
\centering
{\centering \resizebox*{10cm}{7cm}{\rotatebox{-90}{\includegraphics{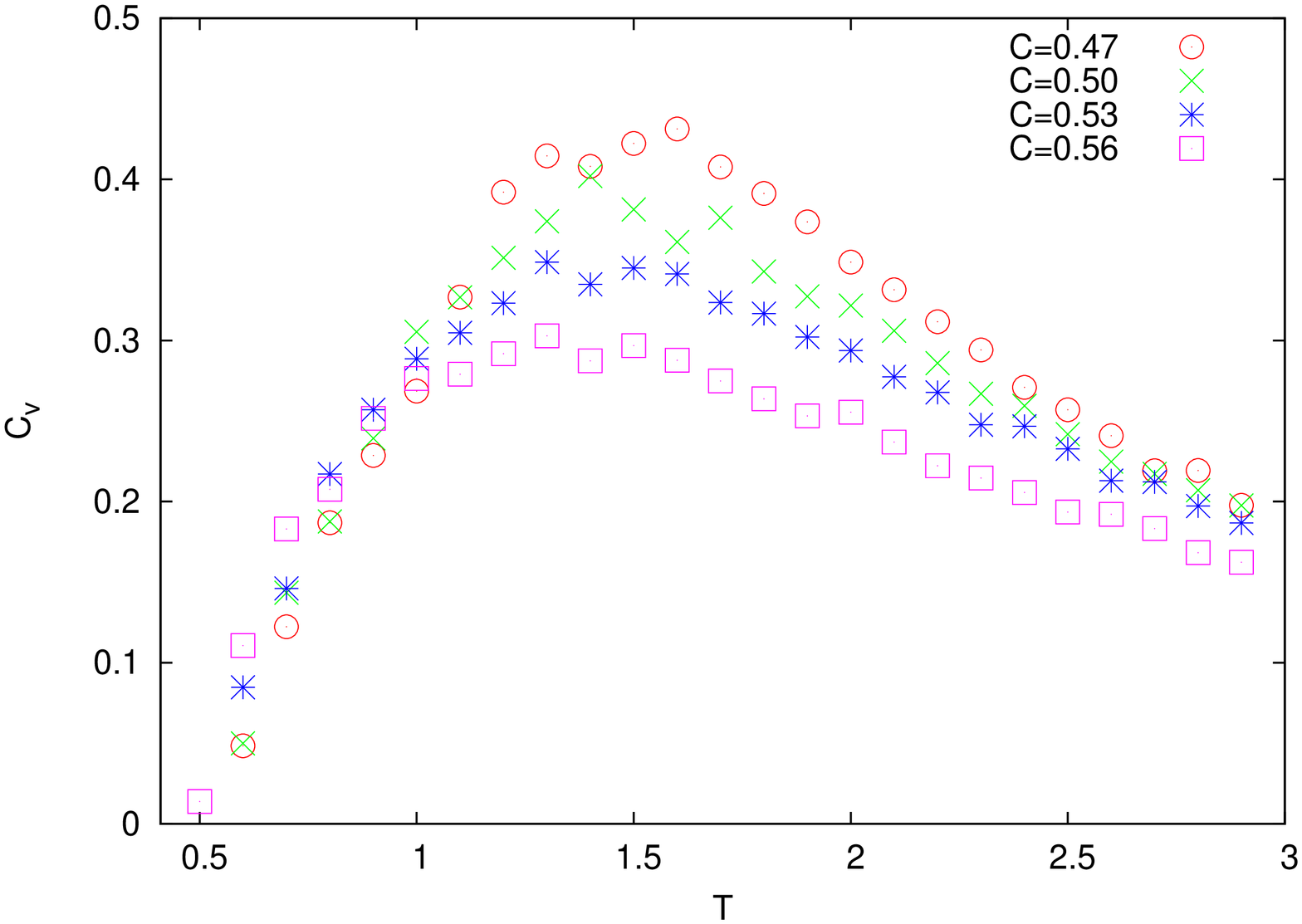}}}} 
\caption{The magnetic specific heat $C_v(T)$ for $<k>=9$ and different values of the clustering coefficient $C$.}
\label{fig-5}
\end{figure}

\section{Results}

In Fig. 2 we show the thermal dependence of the static magnetic susceptibility $\chi(T)$ for the network with mean degree $<k>=4$. Fig. 3 displays the magnetic specific heat $C_v(T)$ for the same network. The plots of the same quantities for $<k>=9$ are shown in Figs. 4 and 5. The positions
of the maxima of $\chi(T)$ and $C_v(T)$ can be treated as approximate values of the transition temperature $T_{SG}$ \cite{mb,by}. Most curves displayed
show some maxima except two cases with highest $C$ for $<k>=4$, where the susceptibility for low temperatures does not decrease - this is shown
in Fig. 2. Still it is clear that the observed maxima do not appear above $T=1.1$ for $<k=4>$ and above $T=1.7$ for $<k>=9$. Moreover, the data 
suggest that when the clustering coefficient $C$ increases, the positions of the maxima of $\chi(T)$ and $C_v(T)$ decrease. This is visible in particular 
for $<k>=9$, in Figs. 4 and 5.\\

\begin{figure}[ht] 
\centering
{\centering \resizebox*{10cm}{7cm}{\rotatebox{-90}{\includegraphics{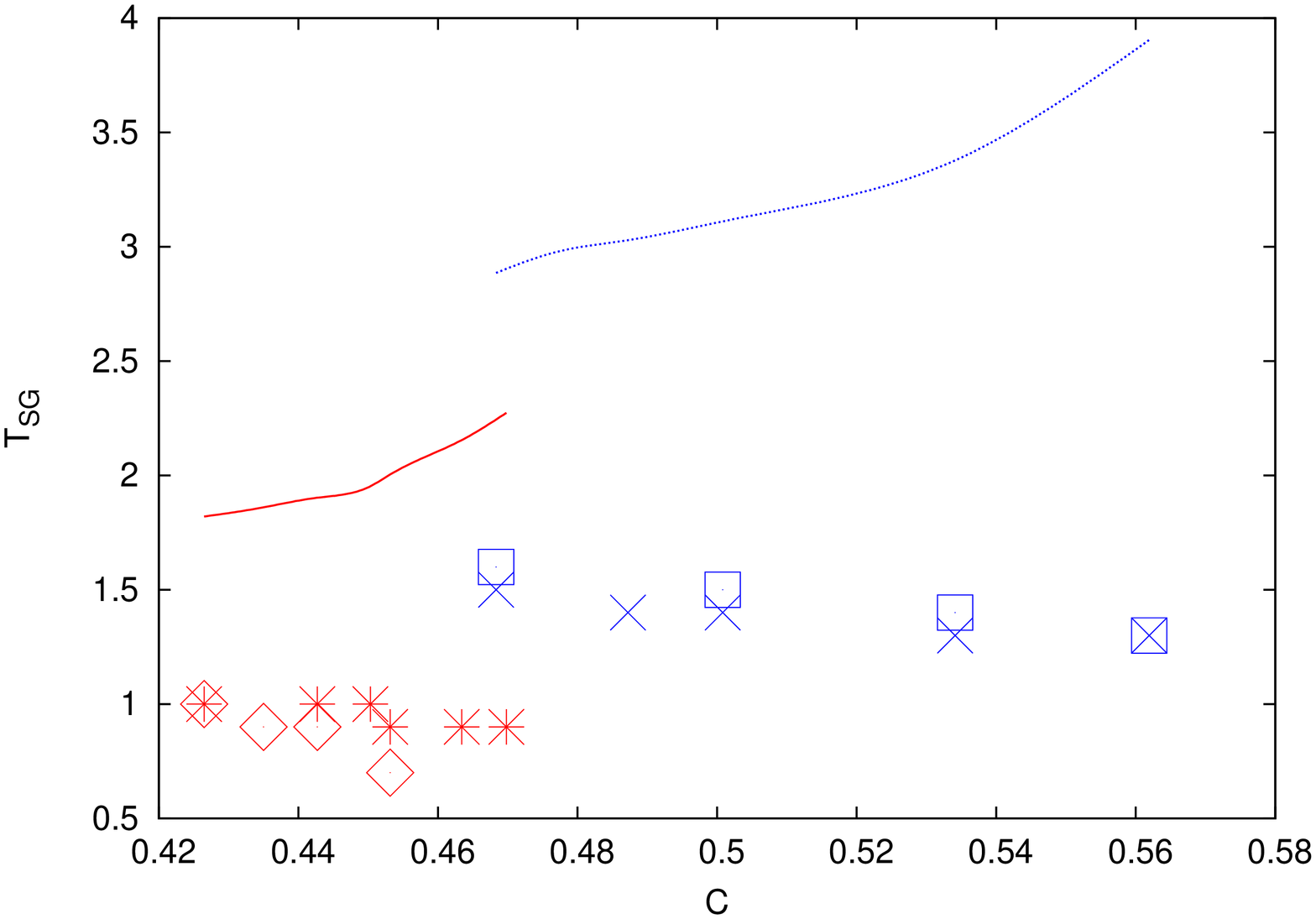}}}} 
\caption{The transition temperature $T_{SG}$ for $<k>=4$ (left side of the figure, in red, continuous line) and $<k>=9$ (right side, in blue, dotted line), against the clustering coefficient $C$. Points mark the results of the numerical simulations: for $<k>=4$ stars come from $\chi$ and rhombs from $C_v$, and for $<k>=9$ X's come from $\chi$ and squares from $C_v$. Lines are the theoretical plots from Eq. 1.}
\label{fig-6}
\end{figure}

In Fig. 6 we show approximate values of the transition temperatures $T_{SG}$, as read from Figs. 2-5. These results are compared with the theoretical values of $T_{SG}$, obtained from Eq. 1. On the contrary to the numerical results, the Bethe theory indicates that $T_{SG}$ is almost constant or increases with $C$. This discrepancy is our main numerical result. It is also of interest that once the clustering coefficient $C$ increases, the susceptibility $\chi$ increases but the specific heat $C_v$ decreases. This can be due with the variation of the shape of the free energy, as dependent on temperature and magnetic field.

\section{Discussion} 

Our numerical results can be summarized as follows. The temperature $T_{SG}$ of the transition from the paramagnetic phase to the spin-glass phase decreases with the clustering coefficient $C$. We interpret this decrease as a consequence of the increase of the density of frustrations. More frustrated
triads make the energy minima more shallow and then a smaller thermal noise is sufficient to throw the system from one to another minimum. This result is in contradiction to the Bethe theory. However, in this theory the frustration effect is neglected. Then the overall picture, obtained previously \cite{amk1,amk2} for the small-world networks, is confirmed here also for the spatial networks.\\

This interpretation can be confronted with recent numerical results of Herrero, where the transition temperature $T_{SG}$ increases
with the clustering coefficient $C$ in the square lattice \cite{hero}. As it is shown in Fig. 7 of \cite{hero}, $T_{SG}$ decreases from 2.3 to 1.7 when the rewiring probability $p$ increases from zero to 0.4. Above $p=0.4$, $T_{SG}$ remains constant or increases very slightly, from 1.7
to 1.72 when $p=1.0$. The overall dependence can be seen as a clear decrease of $T_{SG}$. On the other hand, the clustering coefficient $C$ does not increase remarkably with the rewiring probability $p$. The solution of this puzzle is that in the square lattice with rewiring the frustrations are not due to triads, but to two interpenetrating sublattices, which are antiferromagnetically ordered in the case when $p=0$. The conclusion is that it is the increase of the density of frustrations what always leads to a decrease of $T_{SG}$.\\

A few words can be added on the significance of these results for the science of complexity, with a reference to the computational problem of satisfiability. In many complex systems we deal with a number of external conditions, when all of them cannot be fulfilled. Second premise is that in many complex systems a noise is ubiquitous. These are analogs of frustration and thermal noise. In the presence of noise and contradictive conditions, the system drives in its own way between temporally stable states, similarly to the way how the Ising spin glass wanders between local minima of energy. Once the number of contradictive tendencies or aspirations increases, the overall structure becomes less stable.

\bigskip

{\bf Acknowledgements.} We are grateful to Carlos P. Herrero for his comment. The calculations were performed in the ACK Cyfronet, Cracow, grants No. MNiSW /SGI3700 /AGH /030/ 2007 and MNiSW /SGI3700 /AGH /031/ 2007.

\end{document}